\documentclass[amsmath,amssymb,aps,prb,twocolumn,epsf,floatfix]{revtex4-1}

\usepackage{graphicx}
\usepackage{dcolumn}
\usepackage{bm}
\usepackage{microtype}
\usepackage[dvipsname]{xcolor}
\usepackage{ulem}

\begin{document}

\title{Flavour-selective localization in interacting lattice fermions\\via SU(N) symmetry breaking}

\author{D. Tusi$^1$, L. Franchi$^2$, L. F. Livi$^2$, K. Baumann$^{3,4}$, D. Benedicto Orenes$^{5,6}$, L. Del Re$^7$, R. E. Barfknecht$^5$,\\T. Zhou$^2$, M. Inguscio$^{8,1,5}$, G. Cappellini$^{5,1}$, M. Capone$^{3,9}$, J. Catani$^{5,1}$, L. Fallani$^{2,1,5,10}$}
\affiliation{\mbox{$^1$ LENS European Laboratory for Nonlinear Spectroscopy (Sesto Fiorentino, Italy)}\\
\mbox{$^2$ Department of Physics and Astronomy, University of Florence (Sesto Fiorentino, Italy)}\\
\mbox{$^3$ SISSA Scuola Internazionale Superiore di Studi Avanzati (Trieste, Italy)}\\
\mbox{$^4$ Laboratoire de Physique et Etude des Matériaux, UMR8213 CNRS/ESPCI/UPMC (Paris, France)}\\
\mbox{$^5$ CNR-INO Istituto Nazionale di Ottica del Consiglio Nazionale delle Ricerche (Sesto Fiorentino, Italy)}\\
\mbox{$^6$ ICFO - Institut de Ciències Fotòniques, The Barcelona Institute of Science and Technology (Castelldefels, Spain)}\\
\mbox{$^7$ Department of Physics, Georgetown University, 37th and O Sts., NW, Washington,
DC 20057, USA}\\
\mbox{$^8$ Department of Engineering, Campus Bio-Medico University of Rome (Roma, Italy)}\\
\mbox{$^9$ CNR-IOM Istituto Officina dei Materiali, Consiglio Nazionale delle Ricerche (Trieste, Italy)}\\
\mbox{$^{10}$ INFN National Institute for Nuclear Physics (Firenze, Italy)}\\
}

\begin{abstract}
A large repulsion between particles in a quantum system can lead to their localization, as it happens for the electrons in Mott insulating materials. This paradigm has recently branched out into a new quantum state, the orbital-selective Mott insulator, where electrons in some orbitals are predicted to localize, while others remain itinerant. We provide a direct experimental realization of this phenomenon, that we extend to a more general flavour-selective localization. By using an atom-based quantum simulator, we engineer SU(3) Fermi-Hubbard models breaking their symmetry via a tunable coupling between flavours, observing an enhancement of localization and the emergence of flavour-dependent correlations. Our realization of flavour-selective Mott physics opens the path to the quantum simulation of multicomponent materials, from superconductors to topological insulators.
\end{abstract}

\maketitle

Interactions shape the collective behavior of many-particle quantum systems, leading to rich phase diagrams where conventional and novel  phases can be induced by a controlled variation of external stimuli. The most direct example is perhaps the Mott insulator, where the large repulsion amongst particles leads to an insulating state despite the non-interacting system being a metal or a superfluid\cite{mott}. In solid-state physics the interest in Mott insulators is reinforced by the observation that the proximity to a Mott transition is a horn of plenty where a variety of novel and spectacular phases can be observed, high-T$_c$ superconductivity\cite{pwa,review_lee} being only the tip of an iceberg\cite{iceberg}.

In recent years it has become clear that the standard SU(2) Fermi-Hubbard model is only a specific example, as many interesting materials require a description in terms of ``multicomponent" Hubbard models, e.g. when the conduction electrons have an additional orbital degree of freedom. These systems are not merely more complicated, rather they host new phenomena, challenging the standard paradigm of Mott localization\cite{review_gdm}. Indeed, when the symmetry between orbitals is broken, by some field or internal coupling, electrons in specific orbitals (or some combinations of them) can be Mott-localized while others remain itinerant, leading to surprising ``orbital-selective Mott insulators" \cite{review_vojta}.
\begin{figure}[t!]
	\centering
	\includegraphics[width=\columnwidth]{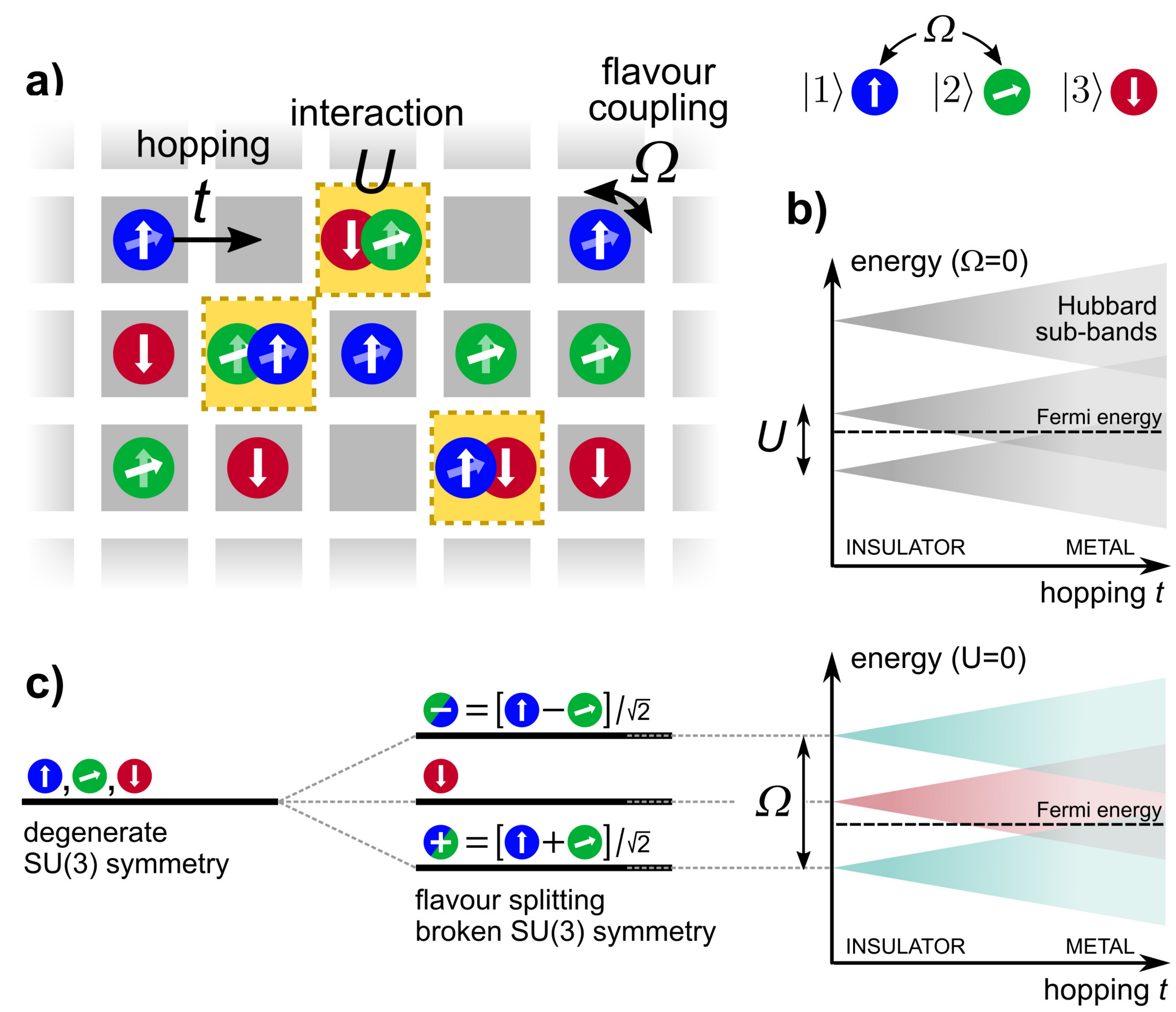}
	\caption{{{\bf Sketch of the physical system.} {\bf a)} We consider a system of repulsively SU(3)-interacting fermions in a lattice. The global symmetry is broken} by a coherent Rabi driving $\Omega$ between two internal states. {\bf b)} In the absence of coupling, the system experiences a phase transition from an SU(3) metal to a Mott insulator as the hopping is reduced. {\bf c)} The Rabi driving lifts the degeneracy between the states and, in a dressed-basis picture, causes them to acquire different energies. The competition with the hopping can drive a transition from a metal to an insulator already in the noninteracting case.}
	\label{fig1}
\end{figure}

\begin{figure*}[t]
	\includegraphics[width=0.95\textwidth]{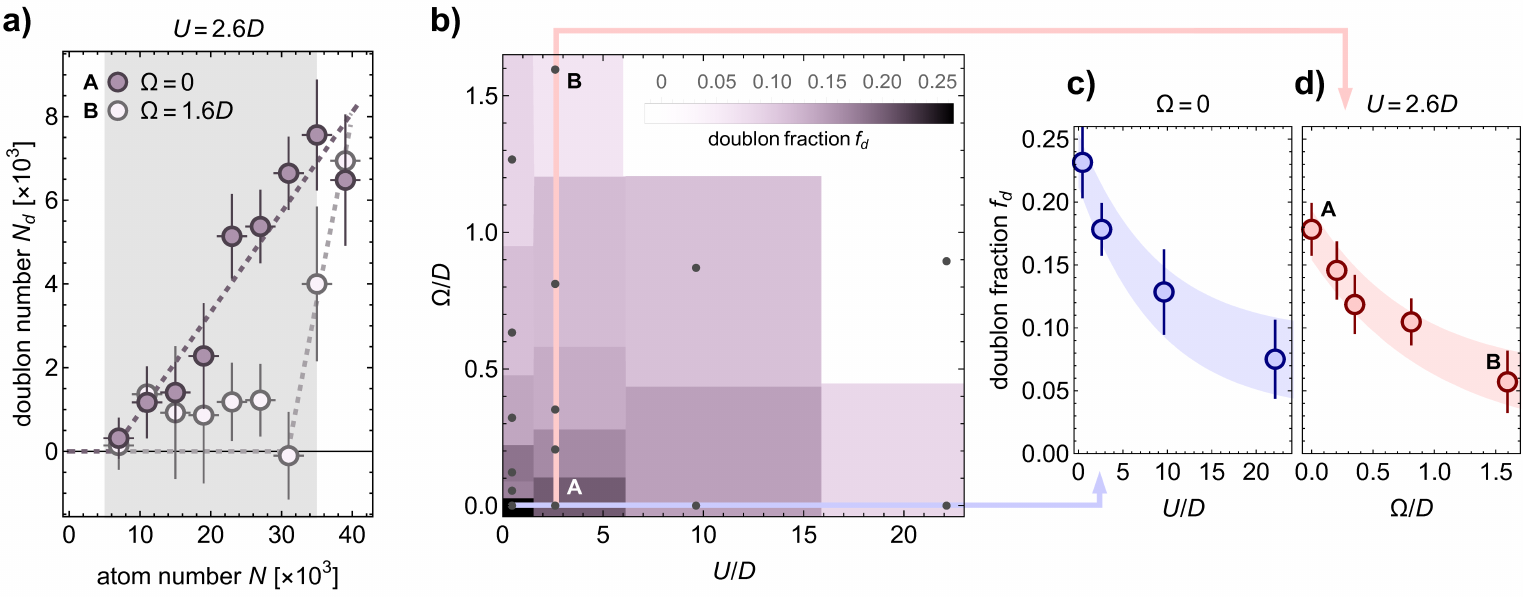}
	\caption{ {\bf Measurement of double occupancies.}  {\bf a)} Number of atoms in doubly-occupied sites $N_d$ as a function of the total atom number $N$ for two different values of $\Omega$ and the same $U=2.6D$. {\bf b)} Average doublon fraction $f_d$ as a function of $U/D$ and $\Omega/D$. The actual measurements are marked by the points. {\bf c,d)} Subsets of the data are shown for two different cross sections of the plot in b), i.e. for $\Omega=0$ (c) and for $U=2.6D$ (d). The measurements in b,c,d) are averages over an interval $N = (5\ldots 35)\times10^3$. Error bars in a,c,d) are obtained with a bootstrap analysis\cite{supplementary}. Dotted lines in a) are fits with a piecewise function (null + straight line), while color shades in c,d) are guides to the eye representing the experimental uncertainty.}
	\label{fig2}
\end{figure*}

Orbital-selective Mott physics has become a central concept for the description of a new class of high-T$_c$ superconductors based on iron\cite{hosono}, as it can describe the anomalies of the metallic state\cite{prl_dgc,natmat_c} and the orbital character of  superconductivity\cite{seamus} in those systems. However, a clean observation of selective Mott transitions is intrinsically hard in solid-state systems, because of the limited experimental control over the microscopic parameters and the orbital degree of freedom. The paradigm of selective Mott physics is itself far from being fully explored and has the potential to become a powerful framework to understand a variety of phenomena, from superconductivity to topological properties, in multicomponent quantum materials.

In this work we take a broader perspective and treat orbital-selective Mott physics as an example of the general concept of ``flavour-selective Mott localization"\cite{delre2018}, which can be realized in a variety of multi-flavour systems, where the flavour can be the spin, the orbital or any other quantum number. We realize a minimal instance of this phenomenon by means of an atomic quantum simulator based on the optical manipulation of nuclear-spin mixtures of ultracold two-electron $^{173}$Yb atoms. This platform allows the realization of multicomponent systems with global SU(N) interaction symmetry \cite{gorshkov2010, cazalilla2014}, as in recent works reporting the realization of SU(N) quantum wires \cite{pagano2014}, SU(N) Mott insulators \cite{taie2012,hofrichter2016} and, more recently, SU(N) quantum magnetism \cite{ozawa2018}. Here we introduce a novel approach to break the symmetry of SU(3) Fermi-Hubbard systems in a controlled way, which allows us to go beyond the investigations in the solid state and to observe directly the two key signatures of selective-Mott physics\cite{delre2018}: an overall enhancement of Mott localization and the onset of flavour-selective correlations.

The experiment is performed with three-component ultracold $^{173}$Yb Fermi gases with total atom number up to $N=4 \times 10^4$ and temperature $T \simeq 0.2 T_F$. The atoms are trapped in a cubic 3D optical lattice (lattice constant $d=\lambda/2=380$ nm), which realizes the multi-flavour Hubbard Hamiltonian

\begin{equation}
\label{eq:hamiltonian}
\begin{aligned}
\hat{H} = &-t \sum_{\langle i,j\rangle,\alpha} \left( \hat{c}_{\alpha i}^\dagger \hat{c}_{\alpha j} +\mathrm{h.c.} \right) +U \sum_{i, \alpha,\beta \neq \alpha} \hat{n}_{\alpha i} \hat{n}_{\beta i}+ \hat{V}_T\\&+\frac{\Omega}{2} \sum_{i} \left(\hat{c}_{1 i}^\dagger \hat{c}_{2 i} +\mathrm{h.c.} \right) 
\end{aligned}
\end{equation}
where $\alpha,\beta \in \{1,2,3\}$ indicate the fermionic flavours (corresponding to nuclear spin states $m=+5/2,+1/2$ and $-5/2$, respectively), $t$ is the tunnelling energy between nearest-neighboring sites $\langle i,j \rangle$, $U$ is the onsite repulsion energy between two atoms of different flavours, and $\hat{V}_T=\kappa \sum_{i,\alpha}R_i^2 \hat{n}_{\alpha i}$ describes the effects of a slowly-varying harmonic trapping potential (where $R_i$ is the distance of site $i$ from the trap center). In the absence of the 4th term of Eq. (\ref{eq:hamiltonian}), $\hat{H}$ has an intrinsic global SU(3) symmetry, which is ensured by the invariance of atom-atom interactions on the spin state and by the realization of spin-independent optical potentials, i.e. $U$, $t$ and $\kappa$ do not depend on $\alpha$. This symmetry is explicitly broken by the 4th term, which describes a coherent on-site coupling between flavours $|1\rangle$ and $|2\rangle$. This coupling is provided by a two-photon Raman process with Rabi frequency $\Omega/h$\cite{supplementary} (where $h$ is the Planck constant). At the single-particle level, this coupling lifts the degeneracy between the  flavours, creating two dressed combinations $|\pm \rangle = \left(  |1\rangle \pm |2\rangle \right) /\sqrt{2}$, energy-shifted from $|3\rangle$ by $\pm \Omega/2$, as sketched in Fig. \ref{fig1}c. 

\begin{figure*}[t]
	\includegraphics[width=0.85\textwidth]{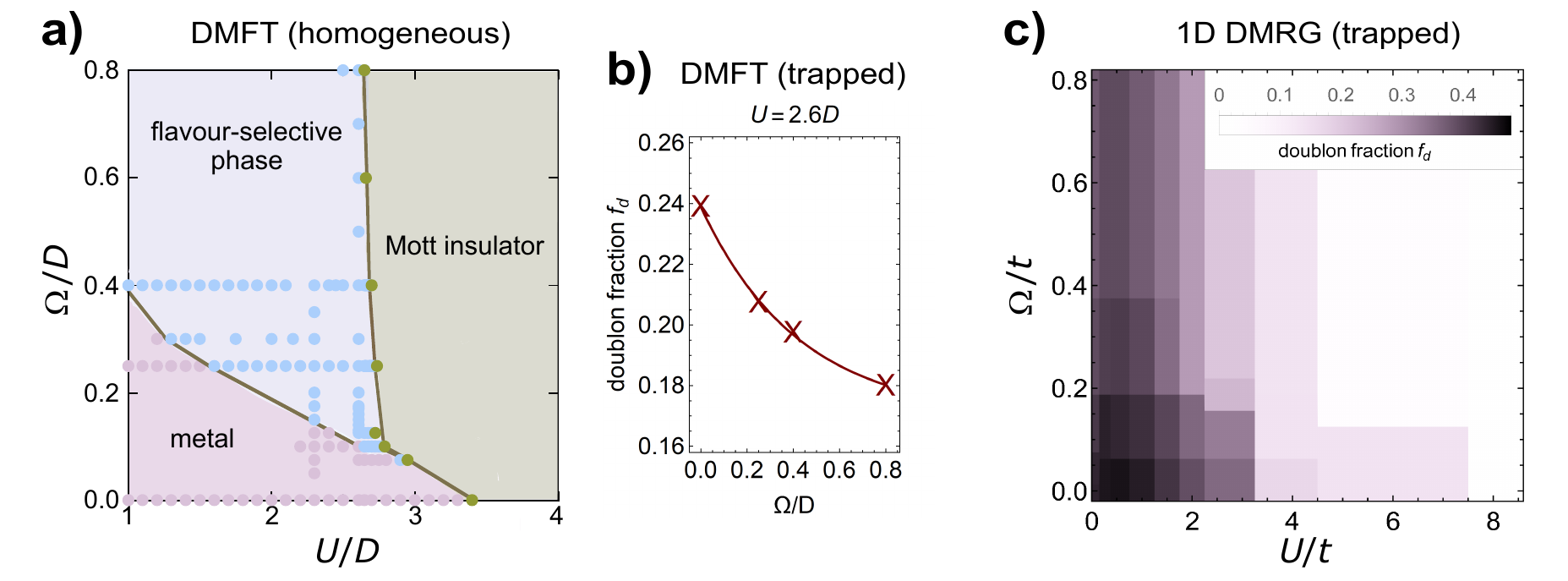}
	\caption{{\bf Theoretical analysis. }{\bf a)} DMFT phase diagram for a homogeneous system at $T=0$ and 1/3 filling. The points refer to different parameter runs. The boundary between the standard metal and the region of selective correlations marks a sharp crossover where the quasiparticle weight of the coupled flavours rapidly drops to a value smaller than 0.05. {\bf b)} Doublon fraction for the trapped system at $U=2.6D$ and different $\Omega/D$, obtained from a LDA analysis of the homogeneous DMFT results with the experimental parameters. {\bf c)} Doublon fraction obtained from DMRG for a 1D trapped system of $N=21$ particles at $T=0$.}
	\label{fig3}
\end{figure*}

We use an adiabatic preparation sequence to produce an equilibrium state of the atomic mixture in the optical lattice, with equal state populations $N_1=N_2=N_3=N/3$ and in the presence of the coherent coupling $\Omega$ between states $|1\rangle$ and $|2\rangle$\cite{supplementary}. In order to characterize the degree of localization of the particles in the lattice, we measure the number of atoms $N_d$ in doubly-occupied sites (called ``doublons" in the following) with photoassociation spectroscopy, as in previous experiments that demonstrated the onset of Mott localization for ultracold fermions \cite{jordens2008,taie2012,supplementary}. In Fig. \ref{fig2}a we report typical measurements of $N_d$ as a function of the total atom number $N$. In a harmonically trapped system, the rate of change of $N_d$ vs $N$ gives information on the core compressibility \cite{jordens2008,jordens2010}. A vanishing value of $N_d$ over an extended range of $N$ signals the presence of an incompressible state with one atom per site in the center of the trap (since adding particles does not lead to a proportional increase of doublons), while the critical $N$ above which $N_d$ then departs from zero can be connected to the magnitude of the energy gap protecting the localized phase. Fig. \ref{fig2}a shows two datasets for $U=2.6D$ (where $D=6t$ is the tunnelling energy times the lattice coordination number), and two different Rabi couplings $\Omega=0$ and $\Omega=1.6D$, respectively. The range of $N$ in the figure spans a range of local fillings up to $\approx 0.8$ atoms/site per state (for the largest $N$) in the center of the trap in the noninteracting case. The comparison between the two datasets clearly shows that the Rabi coupling $\Omega$ results in an enhanced suppression of doublons, enlarging the region of $N$ where the incompressible state forms.

In the following we take the doublon fraction $f_d=\langle N_d/N \rangle$, averaged over the $N$ interval marked by the gray region in Fig. \ref{fig2}a, as an indicator of the degree of Mott localization of the system. The measured values of $f_d$ are shown as a function of $U$ and $\Omega$ in the diagram of Fig. \ref{fig2}b, clearly revealing the cooperative effect of Rabi coupling and repulsive interactions driving the system towards a  Mott-localized state. The same data are plotted with error bars in Figs. \ref{fig2}c,d along two different line cuts of the diagram in Fig. \ref{fig2}b. Fig. \ref{fig2}c shows the effect of an increasing $U$ in the transition towards an SU(3) Mott insulator for $\Omega=0$, while  Fig. \ref{fig2}d shows a similar localization effect induced by $\Omega$ at a fixed interaction strength $U=2.6D$.

We now show the comparison of the experimental results with different theoretical analysis of the model in Eq. (\ref{eq:hamiltonian}). Fig. \ref{fig3}a shows a zero-temperature phase diagram obtained from Dynamical Mean-Field Theory (DMFT)\cite{rev_dmft} for the homogeneous system ($V_T=0$) with a uniform 1/3 filling (one atom per site) and equal number of particles for each flavour. The phase diagram clearly has the same shape of the experimental one, showing that the Hubbard $U$ and the coupling $\Omega$ cooperate in driving the system from a metallic phase to a more localized state. Between the standard metal and the Mott phase, we find a region where the degree of correlations (as measured by the quasiparticle weight) is strongly selective, the coupled flavours being much more localized than the uncoupled one. It is evident that both selective and global Mott localization occur at a smaller $U$ if $\Omega$ is included.

In order to connect this effect to the experimentally measured signal, in Fig. \ref{fig3}b we show the doublon fraction $f_d$ obtained from the homogeneous DMFT results after a local-density approximation (LDA) analysis, to take into account the effect of the harmonic trapping in $V_T$. The reduction of $f_d$ with increasing $\Omega$ is in agreement with the experimental observations reported in Fig. \ref{fig2}. The lack of a quantitative matching with Fig. \ref{fig2}d can be attributed to imperfections in the initial state preparation and to the finite temperature of the experiment, resulting in an average entropy per particle $S/N \approx 2.5 k_B$, which is known\cite{jordens2010} to produce an effect on the double occupancies in a trapped system already at $\Omega=0$.

\begin{figure}[t!]
	\centering
	\includegraphics[width=0.88\columnwidth]{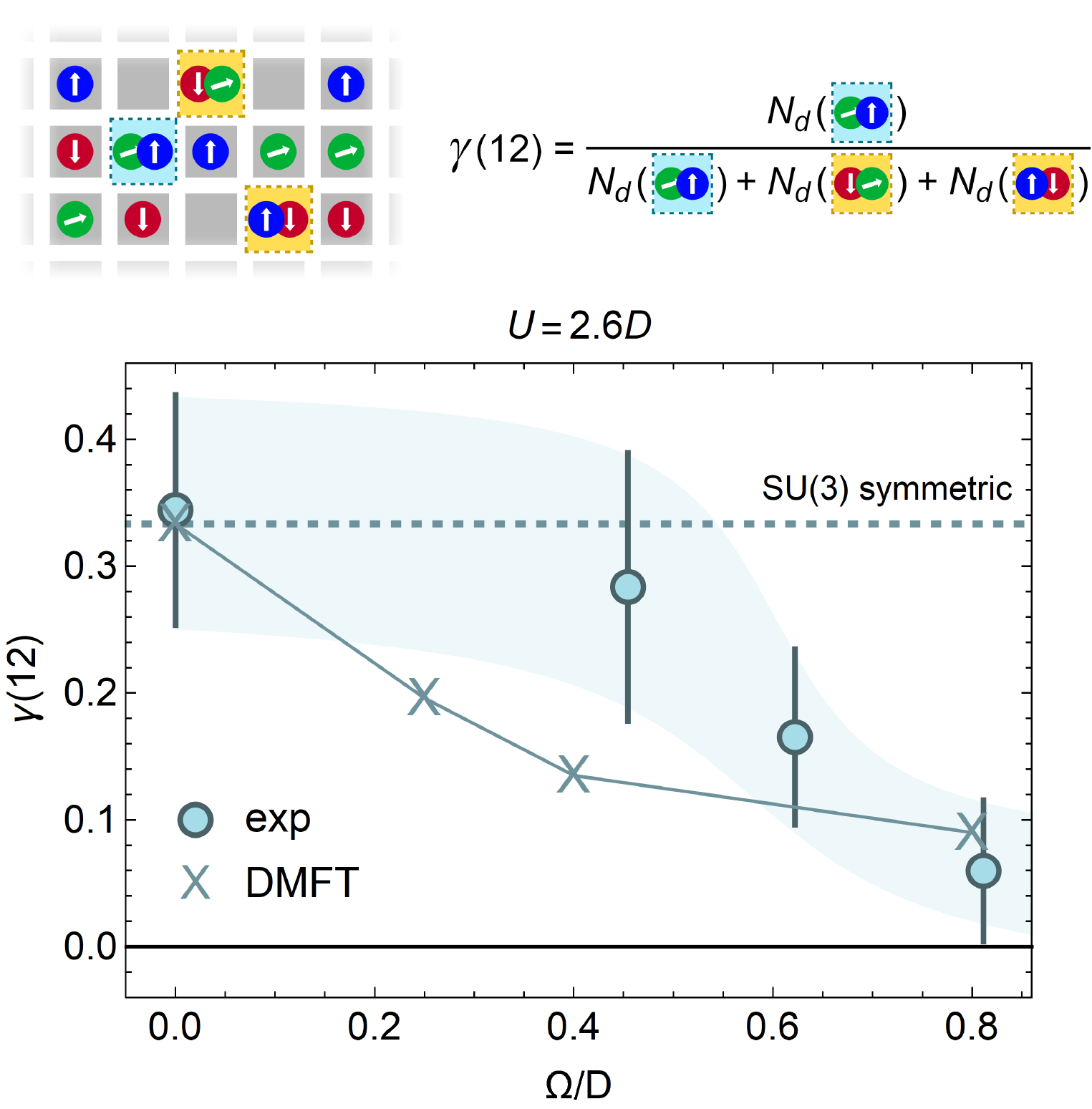}
	\caption{{\bf Evidence of state-selective correlations.} Experimental values of $\gamma(12)$ as a function of $\Omega/D$ for a fixed $U=2.6D$. The circles are averages over an interval $N = (5\ldots 35)\times10^3$, error bars are obtained from a bootstrap analysis\cite{supplementary}, while the color shade is a guide to the eye. The crosses are obtained from a LDA analysis of the DMFT resultes. The dashed line shows the expected value for an SU(3)-symmetric system.}
	\label{fig4}
\end{figure}

In Fig. \ref{fig3}c we finally show the result of a zero-temperature DMRG calculation of $f_d$ for a harmonically trapped 1D system\cite{supplementary}. Although a quantitative agreement with the experimental data should not be seeked (because of the different dimensionality and the finite temperature of the experimental realization), the overall behavior, i.e. a reduction of doublons for increasing $U$ and $\Omega$, is correctly captured and agrees also with DMFT. This also indicates that the phenomena we are exploring are generic and qualitatively independent on the dimensionality.

In addition to the enhancement of Mott localization, the DMFT phase diagram of Fig. \ref{fig3}a shows that $\Omega$ is expected to result in a flavour-dependent localization of the many-body system. The flavour-selective behavior can be experimentally detected by resolving the spin character of the doublons, i.e. counting how many atoms form doublons in each of the three pairs $|12\rangle$, $|23\rangle$, $|31\rangle$ {by means of state-selective photoassociation at a high magnetic field\cite{supplementary}}. In Fig. \ref{fig4} we plot the quantity $\gamma(12)=N_d(12)/N_d$, where $N_d(12)$ is the number of atoms forming doublons in the $|12\rangle$ channel, as a function of $\Omega$ and fixed $U=2.6D$. The measured value at $\Omega=0$ agrees with the expectation for an SU(3)-symmetric system, for which $\gamma(12)=1/3$. As $\Omega$ is increased and the SU(3) symmetry is broken, $\gamma(12)$ diminishes, eventually approaching zero for $\Omega \approx D$. The doublons acquire a strongly flavour-selective behavior.

This suppression of $|12\rangle$ doublons is triggered by the polarization effect in the rotated $|\pm \rangle$ basis, which can be understood, at a qualitative level, already from the simplified, noninteracting case sketched in Fig. \ref{fig1}c: while $|23\rangle$ and $|31\rangle$ doublons can be formed by two fermions in the lowest single-particle states $|+\rangle$ and $|3\rangle$, $|12\rangle$ doublons can be formed only if the two fermions occupy states  $|+\rangle$ and $|-\rangle$, therefore with an additional energy cost $\Omega/2$. Interactions then increase this effect\cite{delre2018}, leading to strong flavour-selective results even for small values of $\Omega$. 

The crosses in Fig. \ref{fig4} are the result of a DMFT calculation in which the flavour populations are kept equal and the harmonic trapping $V_T$ has been taken into account in a LDA approach. The results of this numerical calculation are in good agreement with the experimental findings,  with a larger degree of selectivity for the theoretical calculation. We argue that a better agreement could be seeked by including finite-temperature effects in the calculation: indeed, we expect the state-selective behavior to be reduced by the thermal occupation of higher-energy states, leading to an effective reduction of the polarization in the  $|+\rangle$, $|-\rangle$ basis.

We note that the experimental observation of a finite number of $|12\rangle$ doublons at small, but finite $\Omega$ is an indication of the validity of the protocol used for the preparation of the atomic state, which is different for $\Omega=0$ and $\Omega>0$\cite{supplementary}. We also note that, despite the polarization in the dressed  $|+\rangle$, $|-\rangle$ basis discussed before, we have verified that the populations of the bare states $|1\rangle$, $|2\rangle$, $|3\rangle$ remain always equal under all the experimental conditions we have considered. This is indeed an important aspect of our experiment. If the populations of the bare states were not fixed (and in particular $N_3$ was left free to adapt), the dressed states would be populated according to the scheme in Fig. \ref{fig1}c, leading to $N_1 = N_2 > N_3$, which favours flavour-selective physics already in the non-interacting system. The experimental state preparation procedure {counteracts} the trivial differentiation between flavours by forcing an even occupation. Therefore, we conclude that the flavour selectivity we observed is essentially due to quantum correlations induced by interactions.

In the theoretical DMFT calculations, the equal population constraint is enforced including an external field {$h$}\cite{supplementary} that favours the occupation of $|3\rangle$ in such a way to match the experimental condition $N_1+N_2+N_3=N/3$. Comparing with Ref. \onlinecite{delre2018}, where the populations were left free, we observe that the quantum correlations leading to the selective regime survive to the inclusion of the field {$h$}. In general terms, single-particle effects trigger flavour selectivity, but the inclusion of interactions strongly enhances the differentiation, turning a minor modulation of kinetic energy into a quantitative phenomenon which can also lead to a selective Mott transition\cite{osmt2009}. This is a very general framework, underlying many investigations of multicomponent models.

Exploiting an idealized quantum simulator we have obtained a clear-cut evidence for correlation-induced flavour-selective physics, where the SU(N) symmetry-breaking coupling $\Omega$ is only the trigger of a flavour-selective phenomenon which is fundamentally driven by correlation effects. Our first realization of multicomponent Hubbard physics with coherent internal couplings opens new paths for the quantum simulation of new classes of materials ranging from high-temperature superconductors to interacting topological insulators as described by the Bernevig-Hughes-Zhang model \cite{BHZ2006}. As the coupling is realized with a nonzero momentum transfer (i.e. non collinear Raman beams), a full range of possibilities will emerge, including the study of magnetic crystals \cite{barbarino2015}, fractional quantum Hall states \cite{calvanese2017}, and the effect of interactions of topological phase transitions\cite{amaricci2015} and on the associated edge states\cite{amaricci2017}.

{\bf Acknowledgments.} We acknowledge insightful discussions with M. Dalmonte, D. Cl\'{e}ment, F. Scazza and financial support from projects TOPSIM ERC Consolidator Grant, TOPSPACE MIUR FARE project, QTFLAG QuantERA ERA-NET Cofund in Quantum Technologies, MIUR PRIN project 2017E44HRF, MIUR PRIN project 2015C5SEJJ, MIUR PRIN project 20172H2SC4\_004, INFN FISh project.
L.D.R was supported by the U.S. Department of Energy, Office of Science, Basic Energy Sciences, Division of Materials Sciences and Engineering under Grant No. DE-SC0019469.


\newpage


\renewcommand{\thefigure}{S\arabic{figure}}
 \setcounter{figure}{0}
\renewcommand{\theequation}{S.\arabic{equation}}
 \setcounter{equation}{0}
 \renewcommand{\thesection}{S.\Roman{section}}
\setcounter{section}{0}
\renewcommand{\thetable}{S\arabic{table}}
 \setcounter{table}{0}

\onecolumngrid
\newpage
\clearpage
\newpage

\begin{center}
{\bf \large Supplementary Material for\\
\vspace{3mm}
``Flavour-selective localization in interacting lattice fermions\\via SU(N) symmetry breaking"}\\
\vspace{3mm}

D. Tusi, L. Franchi, L. F. Livi, K. Baumann, D. Benedicto Orenes, L. Del Re, R. E. Barfknecht\\
T. Zhou, M. Inguscio, G. Cappellini, M. Capone, J. Catani, L. Fallani
\end{center}

\vspace{6mm}
\twocolumngrid

\section{Measurement of double occupancies}
\label{sec:PA}

We measure double occupancies (doublons) in the 3D lattice exploiting a one-color photoassociation (PA) process which transfers the atoms in doubly-occupied lattice sites into highly excited molecular states \cite{jordens2008}. The photoassociated atoms are rapidly lost from the lattice due to the fast decay of the molecular states, leaving a sample consisting only of singly-occupied sites. The number of double occupancies is thus inferred by  first measuring the total number of atoms in the sample and then subtracting the number of atoms in singly occupied lattice sites remaining after the PA pulse. For the regime of atom number considered in this work we expect the number of triply-occupied sites to be negligible. 

Our measurement procedure starts with a rapid freeze of the atomic density distribution that we realize increasing the lattice depth up to $V_0=25 E_{rec}$ ($E_{rec}=h^2/8md^2$, where $h$ is the Planck constant, $m$ the atomic mass and $d$ the lattice spacing) in a time 1.5 ms. At the final lattice depth tunnelling between lattice sites is completely negligible. Two different schemes are then employed for the PA excitation depending on the specific measurement that we want to perform. 

\begin{figure}
	\includegraphics[width=\linewidth]{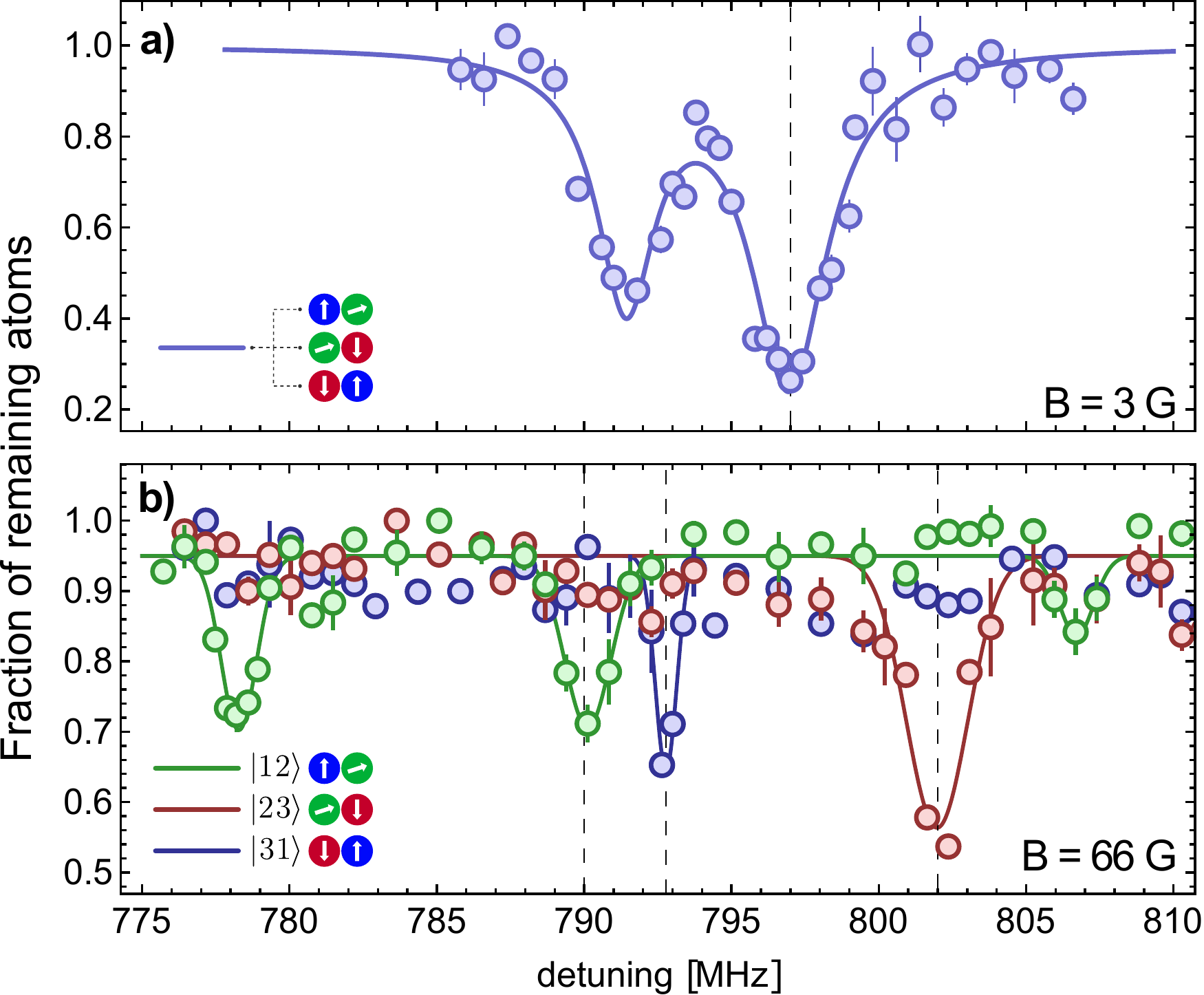}
	\caption{{\bf Detection of double occupancies.} Photoassociation spectroscopy on an SU(3) atomic sample (a) and on different SU(2) mixtures (b). The plots report the number of atoms remaining in the sample after a 5 ms long PA pulse at low laser intensity (unsaturated regime).}
	\label{fig:photoassociation}
\end{figure}

In order to measure the total number of doublons independently on their spin composition (results of Fig. \ref{fig2} of the main text), a 5 ms long, $\sigma^-$ polarized PA pulse is shone on the atomic sample with an intensity $I_{PA}=90$ mW/mm$^2$. The PA pulse excites a molecular line that is 796.2 MHz red-detuned with respect to the ${^1S_0}(F=5/2) \rightarrow {^3P_1}(F=7/2)$ atomic transition \cite{taie2012}. During the process, atoms are subjected to a low magnetic field $B$ = 3 G which defines a quantization axis for the spin, but is low enough not to unveil the Zeeman substructure of the molecular line, as it is shown in the spectrum reported in Fig. \ref{fig:photoassociation}a.

In the presence of a higher external magnetic field, the PA transition frequency is shifted accordingly to the projection $M_T$ of the molecular total angular momentum on the quantization axis \cite{Han2018}. $M_T$ is a quantum number conserved during the PA process, with a value given by the relation $M_T=m_1+m _2+\sigma$, where $m_{1,2}$ are the nuclear spin projections of the the two atoms composing the molecule and $\sigma=-1$ is the angular momentum transferred by the PA photon. As long as the sum $m_1+m_2$ unambiguously characterizes an atomic pair, the molecular Zeeman substructure can thus be employed as a tool to distinguish between doublons with different spin flavours.

Taking this into consideration, in order to detect only doublons with a particular spin composition $\{m_1,m_2\}$, we apply a magnetic field $B=66$ G, which induces a Zeeman shift of the order of several MHz between molecular lines with different $M_T$ projection. 
In this case the spectrum is complicated by the presence of a plethora of features originated by the Zeeman splitting of different transitions, which makes it difficult to label individual lines. The association of the PA lines to a particular spin flavour has been realized preparing different SU(2) mixtures and acquiring individual spectra for each of them, as it is shown in Fig. \ref{fig:photoassociation}b. In particular, at this magnetic field, we identify three strong PA lines at 778.3 MHz, 792.8 MHz and 802.0 MHz which are associated to doublons with spin compositions $|12\rangle=\{+5/2, +1/2\}$, $|23\rangle=\{+1/2, -5/2\}$ and $|31\rangle=\{-5/2, +5/2\}$, respectively.

For the measurements reported in Fig. \ref{fig4} of the main text, $\gamma(12)$ is calculated from a spin-resolved measurement of doublons in the $|12\rangle$ and $|31\rangle$ channels as $\gamma(12)=N_d(12)/(N_d(12)+2N_d(31))$, taking into account $N_d(23)=N_d(31)$. This assumption is justified by the equal population of the two flavours $|1\rangle$ and $|2\rangle$ after the adiabatic preparation sequence described in Sec. \ref{sec:adiabaticpreparation}, as shown in Fig. \ref{fig:dressedstatestability}.

\section{Data analysis}

The number of double occupancies is obtained from a direct measurements of the total number of atoms, with and without photoassociation (PA), obtained through standard time-of-flight absorption imaging (see Sec. \ref{sec:PA}). Measurements with and without PA are alternated in time and the number of double occupancies $N_d$ is obtained by the difference between close pairs, in order to compensate for fluctuations in the initial number of atoms $N$ (without PA) due to slow environmental drifts. For the same set of parameters $\left\{U/D,\Omega/D\right\}$ we acquire from 40 to 80 $(N,N_d)$ pairs. 

The experimental points in Figs. \ref{fig2} and \ref{fig4} are obtained by a bootstrapping method in which the raw data are resampled uniformly over a chosen subrange of $N$, in bins of width $\delta N = 4\times10^3$ for Fig. \ref{fig2}a, and over an extended range $N=(5\ldots 35)\times10^3$ for Fig. \ref{fig2}b,c and Fig. \ref{fig4}. The error bars are obtained as the standard deviation of the mean values for different resamplings of the same $N$ interval.

\section{Unsupervised machine-learning analysis}

In addition to the data analysis presented in the main text and detailed in S.II, we have performed an additional analysis of the flavour-selective double occupancies, based on an unsupervised machine-learning approach. The raw datasets  are large, with at least 200 acquisitions of pairs $(N, N_d)$ for each set of experimental parameters. Following the well-established  \textit{K-Means method} \cite{andrewNG}, we consider the first component of each pair, i.e. the number of atoms $N$, and we set (as the only constraint) the number $K$ of groups to divide them into. Once the groups (clusters) are defined, the mean values for both the number of atoms $N$ and the number of doublons $N_d$ for each cluster are evaluated. The error bars are estimated with a bootstrapping procedure. 

\begin{figure}[t!]
\begin{centering}
\includegraphics[width=0.46\textwidth
]{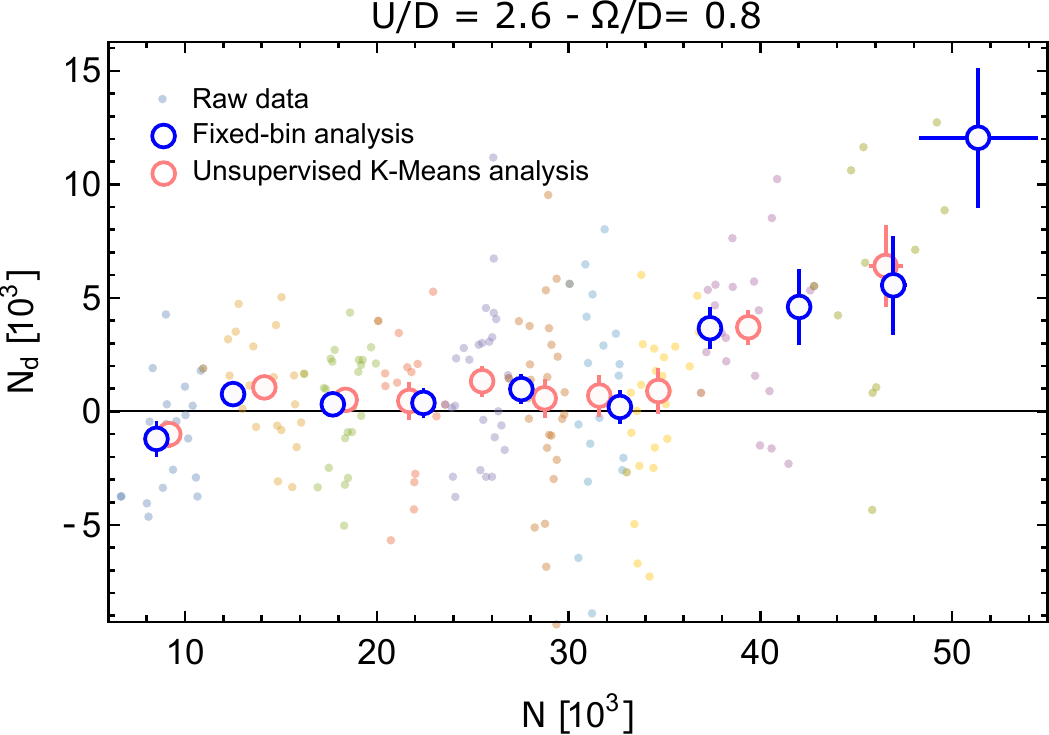}
\caption{\textbf{ Comparison between different data analysis methods.} The number of doublons in the $|12\rangle$ channel is shown as a function of the total atom number $N$ for $U/D = 2.6$ and $\Omega/D = 0.8$, for different data analysis methods.
Small points are the raw $(N,N_d)$ data, with different colours representing the clusters of attribution after the application of the unsupervised K-Means method with $K = 10$ groups. Pink points represent the mean values for the different K-Means clusters. Blue points are obtained after the analysis explained in S.II with fixed bin widths.} 
\label{fig:unComp}
\end{centering}
\end{figure}

In Fig. \ref{fig:unComp} we compare the points obtained with the analysis presented in the main text and those given by the unsupervised K-Means analysis, taking the same number of clusters. The good agreement between them ensures that the specific choice of data analysis procedure does not introduced biases on the results. We note that the K-Means method implies only a constraint on the number of clusters, allowing for a different width among them, differently from the main analysis procedure, where the bin size was fixed. 

\begin{figure}[b]
\begin{centering}
\includegraphics[width=0.47\textwidth 
]{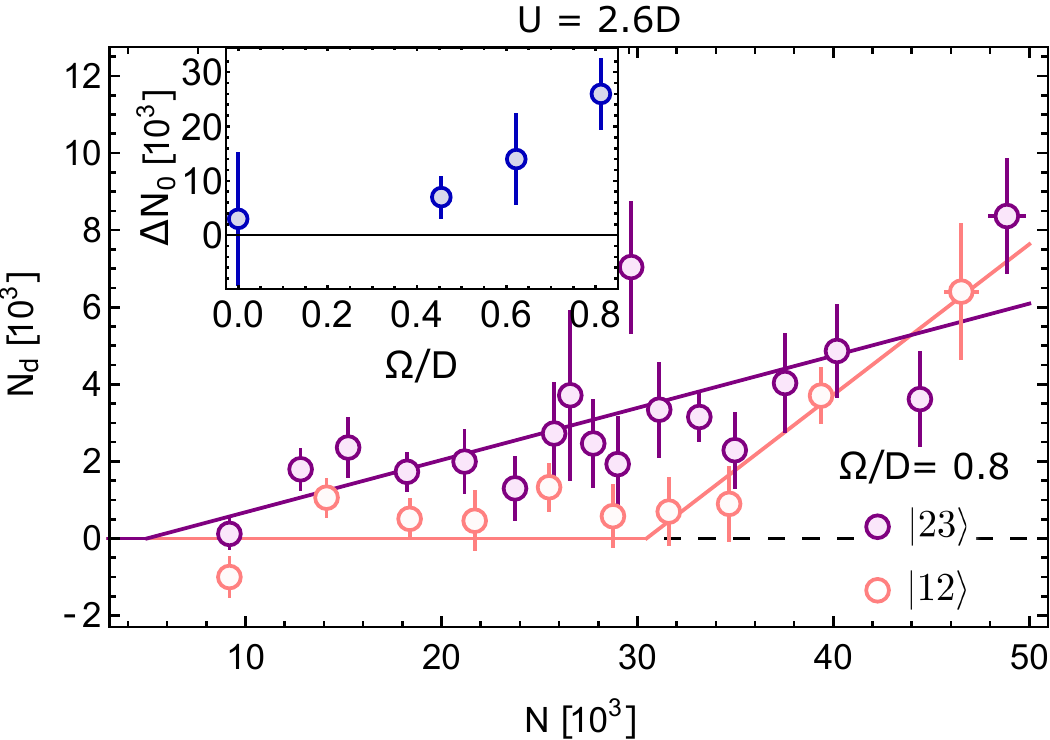}
\caption{\textbf{Flavour-selective localization.} The number of doublons in different flavour channels is shown as a function of $N$ for $U/D = 2.6$ and $\Omega/D = 0.8$, after a K-Means analysis with $K = 18$ clusters. Pink points show the average number of doublons formed by the coupled flavours $\lvert 1 2 \rangle$, while purple points refer to the uncoupled flavours $\lvert 2 3 \rangle$. Solid lines are the fits performed with the piecewise function in Eq. \ref{eq:piecewise}. In the inset we show the difference between the fitted thresholds $\Delta N_0 = N_0(12)-N_0(23)$ as a function of the coupling strength $\Omega$.}
\label{fig:piecewise}
\end{centering}
\end{figure}

As the datasets for the flavour-selective double occupancies are rather large, we can increase the number of groups $K$ to pinpoint the specific range of atom numbers above which doublons form. The flavour-selective averages obtained with the K-Means analysis are fitted with the piecewise function:
\begin{equation}\label{eq:piecewise}
    N_d=
    \begin{cases}
    0 & N \leq N_0 \\
    A (N-N_0) & N>N_0
    \end{cases}
\end{equation}
where $N_0$ defines the threshold atom number below which atoms are localised. In Fig. \ref{fig:piecewise} we show the K-Means averages for the coupled flavours $\lvert 1 2 \rangle$ and the uncoupled ones $\lvert 2 3 \rangle$, together with the results of the fit. The difference between the fitted thresholds for the two doublon flavours $\Delta N_0 = N_0(12)-N_0(23)$ is shown in the inset of Fig. \ref{fig:piecewise} for different values of $\Omega/D$.

This analysis demonstrates that, by increasing the Raman coupling intensity (i.e. lifting the flavour degeneracy more), the threshold for $|12\rangle$ doublons moves to a higher number of atoms, showing an increased localisation in the centre of the trap for the coupled flavours. This result, derived from the unsupervised machine-learning analysis, is consistent with the measurements reported in Fig. \ref{fig4}, and provides another strong indication in support of flavour-selective localization for the symmetry-broken SU(3) Hubbard Hamiltonian.

\section{Implementation of Raman coupling}

The coherent coupling between states $|1\rangle$ ($m=5/2$) and $|2\rangle$ ($m=1/2$) is realized by exploiting a two-photon $\sigma^+ / \sigma^-$ Raman transition. The Raman coupling is implemented with two co-propagating $\lambda = 556$ nm laser beams characterized respectively by  angular frequencies $\omega$ and $\omega+\delta\omega$. In order to reduce the inelastic photon scattering rate, the two beams are 1.754 GHz blue-detuned with respect to the $^1S_0$ $\rightarrow$ ${^3P_1}\,(F=7/2)$ intercombination transition. A 150 G magnetic field is used to define a quantization axis and to remove the degeneracy between the six states of the $^{173}$Yb ground-state manifold, which are split according to their nuclear spin $m$ by $207 \times m$ Hz/G. 

The $\sigma^+/\sigma^-$ coupling between $m=+5/2$ and $m=+1/2$ is obtained by setting the polarization of the two beams to be orthogonal with respect to the quantization axis and adjusting the frequency difference $\delta\omega/2\pi$ in order to compensate the Zeeman splitting and the residual Raman light shift between the two states. As explained in detail in the supplementary material of 
Ref. \onlinecite{Mancini2015}, this choice for the polarization makes the Raman light shifts largely spin-dependent, thus making the $m=+1/2 \leftrightarrow m=-3/2$ coupling nonresonant and effectively excluding state $m=-3/2$ from the dynamics. For a similar reason, also the $m=-5/2 \leftrightarrow m=-1/2$ coupling is nonresonant and state $|3\rangle$ ($m=-5/2$) does not participate to the Raman dynamics.

On the basis of the discussion above, the Raman coupling on the basis formed by the three states $\left\{|1\rangle, |2\rangle, |3\rangle \right\}$ can be described by the following 3x3 rotating-wave-approximated Hamiltonian
\begin{equation} \label{Raman_Hamiltonian}
\hat{H}_R=\frac{\hbar}{2}
\begin{pmatrix} 
\delta & \Omega & 0\\
\Omega& -\delta & 0\\
0 & 0 & 0
\end{pmatrix}
\end{equation}
where $\Omega$ is the angular Rabi frequency associated to the Raman coupling and $\delta$ is the two-photon detuning with respect to the resonance frequency for the $|1\rangle \leftrightarrow |2\rangle$ transition. No detuning is indicated for state $|3\rangle$ as it is decoupled from the Raman process. In the resonant case $\delta=0$, $\hat{H}_R$ corresponds to the third term in the many-body Hubbard Hamiltonian in Eq. (\ref{eq:hamiltonian}) of the main text.

\begin{figure}
	\includegraphics[width=\linewidth]{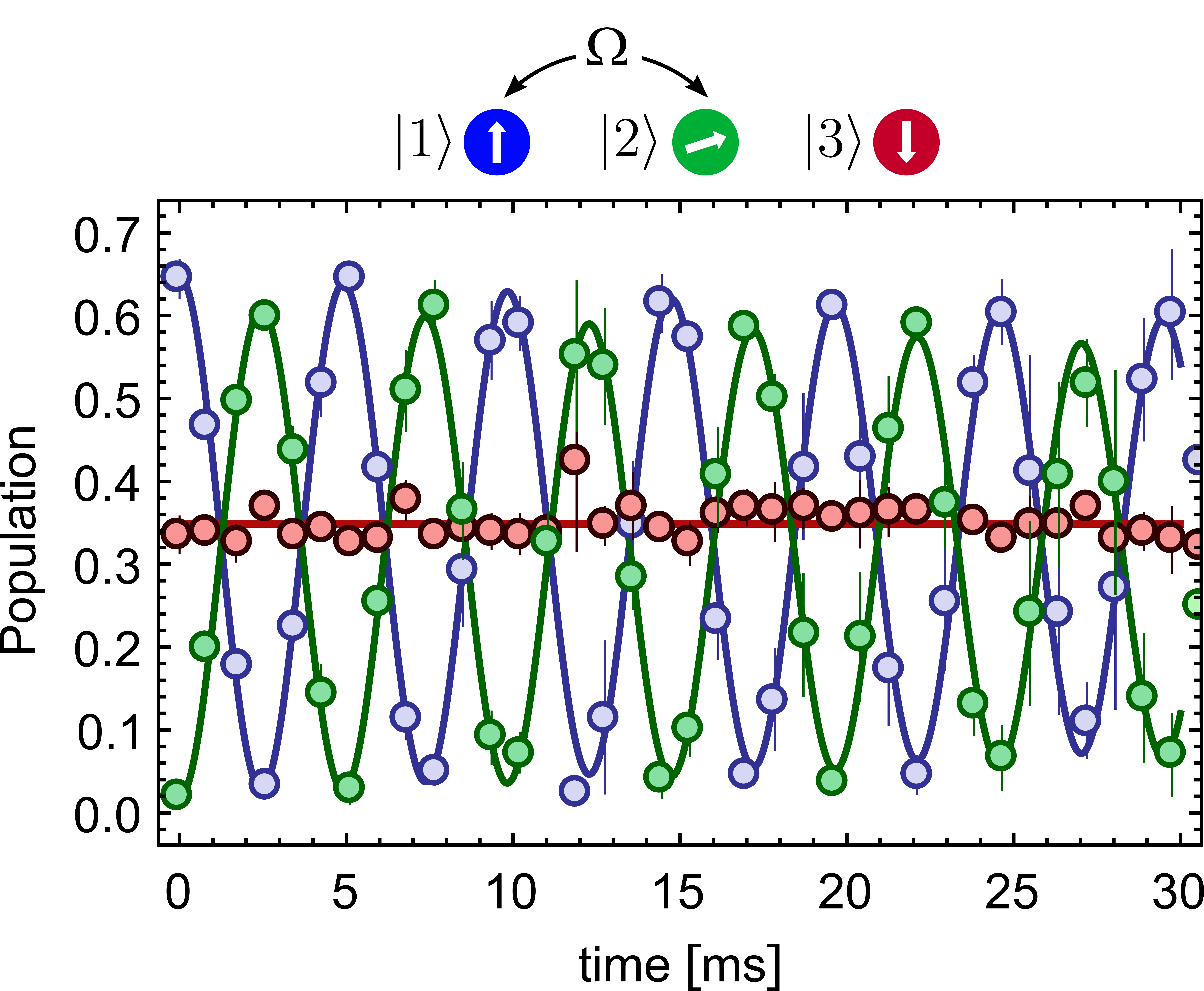}
	\caption{{\bf Coherent Rabi dynamics.} Population dynamics induced by the sudden activation of a resonant Raman coupling between states  $|1\rangle$ and $|2\rangle$ for localized particles in a deep optical lattice (no tunnelling). The points show the fractional population of the states $|1\rangle$ (blue), $|2\rangle$ (green) and $|3\rangle$ (red) as a function of time. The resonant Raman coupling is switched on at time $t=0$, when the initial fractional populations for the three states are $2/3$, $0$ and $1/3$, respectively (as in the adiabatic state preparation procedure described in Sec. \ref{sec:adiabaticpreparation}). The solid lines are sinusoidal fits of the experimental data.}
	\label{fig:raman}
\end{figure}

The Rabi frequency $\Omega$ can be experimentally changed by adjusting the intensities $I_1$ and $I_2$ of the two Raman beams, according to the relation $\Omega \propto \sqrt{I_1 I_2}$. In order to assess the value of $\Omega$ for given values of the Raman beam intensities $I_1$ and $I_2$, we induce resonant Rabi oscillations between states $|1\rangle$ and $|2\rangle$ and determine $\Omega$ from a sinusoidal fit of the experimental data, as shown in Fig. \ref{fig:raman}.

\section{Adiabatic state preparation}
\label{sec:adiabaticpreparation}

A degenerate Fermi gas of $^{173}$Yb atoms, initially confined in a crossed optical dipole trap with harmonic trapping frequencies $\omega_{x,y,z} = 2\pi\times\{55, 96, 73\}$ Hz, is transferred in a simple-cubic 3D optical lattice using a two-step 3s-long adiabatic loading procedure \cite{Mikio2017}. The optical lattice is described by a potential energy $V(x,y,z)=V_0 \left[ \sin^2(\pi x/d)+\sin^2(\pi y/d)+\sin^2(\pi z/d)\right]$, where $d=\lambda/2$ is the lattice spacing and $V_0$ is the lattice energy depth along each of the three principal axes. In the first 2 seconds, the lattice intensity is ramped up with a first spline ramp from $V_0=0$ to $V_0=4 E_{rec}$ ($E_{rec}=h^2/8md^2$ is the recoil energy). After this first step, the lattice intensity is further increased with a 1s-long spline ramp to the final value $V_0$ ranging from $4E_{rec}$ to $15E_{rec}$. During the lattice loading, the depth of the crossed dipole trap is progressively reduced in such a way to keep the harmonic trapping frequencies constant along the three principal axes, independently from the value of $V_0$. 

For the measurements at $\Omega=0$ the loading procedure described above is applied to a balanced 3-component mixture of $^{173}$Yb atoms in the three spin states $|1\rangle$, $|2\rangle$ and $|3\rangle$. The mixture is prepared before the lattice ramp-up procedure with a sequence of optical pumping pulses on the $^1S_0$ $\rightarrow$ $^3P_1$ transition, following the techniques discussed in Ref. \onlinecite{pagano2014}. The populations of the three spin states are all equal, $N_1=N_2=N_3=N/3$, with experimental imperfections on the order of a few $\%$ at most.

For the measurements at $\Omega \ne 0$ the loading procedure starts with a 2-component unbalanced mixture of atoms in states $|1\rangle$ and $|3\rangle$. With a proper choice of optical pumping pulses we adjust the initial populations to be $N_1=2N/3$ and $N_3=N/3$. After the lattice loading, we switch on the Raman beams, initially far detuned from any two-photon transition, and perform an adiatic frequency sweep to bring them resonant with the  $|1\rangle \leftrightarrow |2\rangle$ transition. The resonant condition is reached by means of an exponential frequency sweep of the form
\begin{equation} \label{Raman_detuning}
\delta(t)=\delta_{0} ( e^{-t/\tau} - e^{-T/\tau})/(1 - e^{-T/\tau})
\end{equation}
that reduces the two-photon detuning from $\delta_0 \gg \Omega$ to $\delta=0$ in a time $T$ and time constant $\tau$ (see an example in Fig. \ref{fig:adiabloading}a). This procedure is an adiabatic passage that brings an atom in state $|1\rangle$ to the lowest-energy dressed state of the Raman-coupled system $|+\rangle = \left(|1\rangle+|2\rangle\right)/\sqrt{2}$ (see also the sketch in Fig. \ref{fig1} of the main text). We note that, due to the initial spin unbalance and because of the equal-weighted admixture of states $|1\rangle$ and $|2\rangle$ in the Raman-dressed states, at the end of the detuning ramp the population is equally distributed among the spins, $N_1=N_2=N_3=N/3$, as it was natively in the loading protocol employed for $\Omega=0$.

\begin{figure}[b!]
\includegraphics[width= \linewidth]{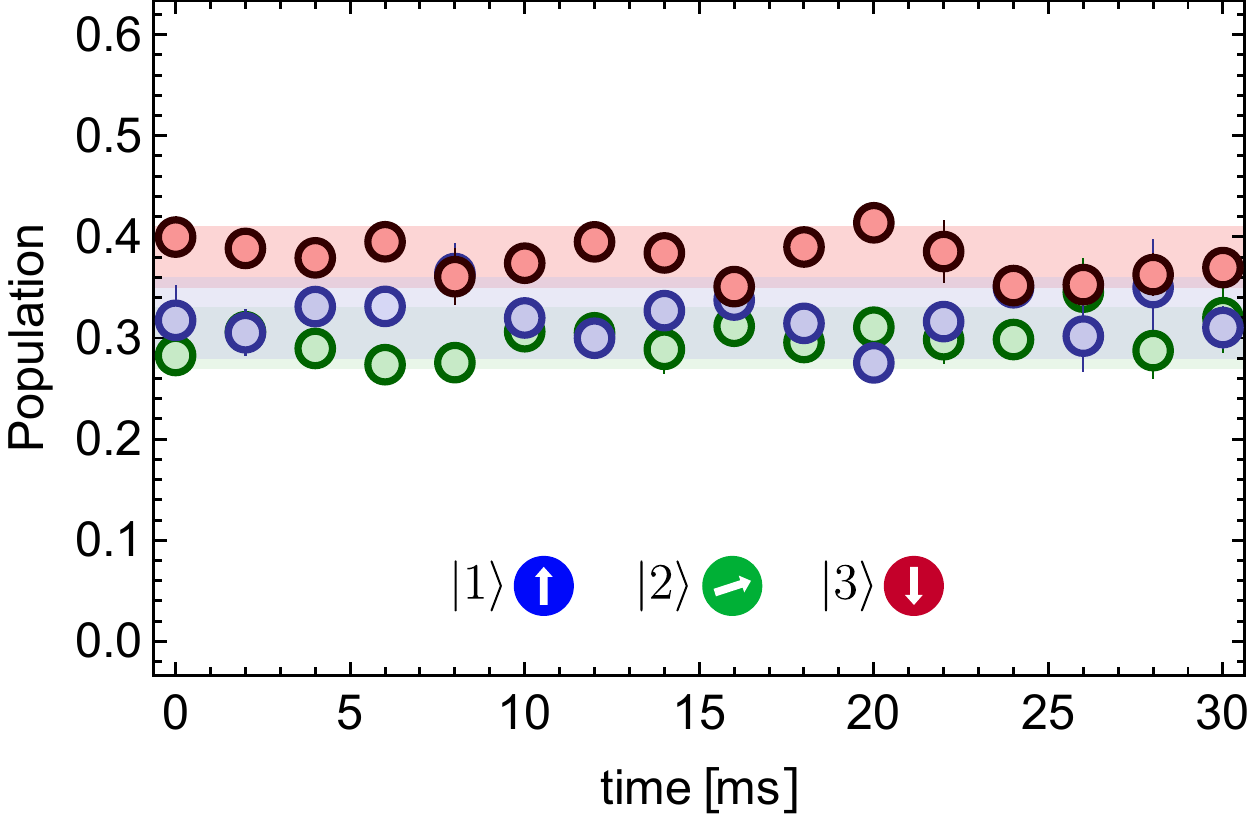}
	\caption{{\bf State population after adiabatic state preparation}. Fractional state population as a function of time after the adiabatic state preparation sequence. Points are the mean value of 3 independent measurements. Shaded regions represent the standard deviation with respect to the mean value. The slight excess of $|3\rangle$ atoms in this dataset has to be attributed to an imperfect optical pumping before the adiabatic state preparation procedure.}
	\label{fig:dressedstatestability}
\end{figure}

The ramp parameters $\delta_0$,  $T$ and $\tau$ are chosen according to a numerical analysis in which we solve the time-dependent Schr\"odinger equation associated to the Raman Hamiltonian in Eq. (\ref{Raman_Hamiltonian}), verifying that at the end of the ramp the initial state is effectively projected onto the Hamiltonian lowest energy eigenstate. Experimentally, we check the adiabaticity of this procedure by verifying that the initial unbalanced  $|1\rangle$-$|3\rangle$ mixture can be recovered with a reversed detuning ramp following that in Eq. (\ref{Raman_detuning}). To further assess the loading fidelity, we verify the time stability of the spin populations at the end of the loading ramp, as shown in Fig. \ref{fig:dressedstatestability}. In both the cases, we measure population differences of a few $\%$ at most, validating the effectiveness of this loading procedure.

\section{Numerical validation of the loading procedure}

In order to assess the validity of the adiabatic state preparation at the many-body level, we have developed a numerical simulation based on the exact diagonalization of Eq. (\ref{eq:hamiltonian}) on a reduced-scale version of our system. We consider $N$ three-component fermions in a 1D lattice with $N_s$ sites, retaining all the relevant terms of Eq. (\ref{eq:hamiltonian}): hopping, repulsive interactions, Raman coupling between states  $|1\rangle$ and $|2\rangle$. We work in the occupation number basis in which the Hilbert space is constituted by the Fock vectors
\begin{equation} \label{Fock_vector}
|\psi\rangle = |n_{1,1},...,n_{\alpha,i},...,n_{3,N_s}\rangle
\end{equation}
where $n_{\alpha,n} =\{0,1\}$ is the occupation number for a particle in internal state $\alpha=\{1,2,3\}$ in the lattice site $i=\{1,...,N_s\}$. In this basis we can write the many-body Hamiltonian as 
\begin{equation} \label{toy_model_hamiltonian}
\begin{aligned}
\hat{H} = & -t \sum_{i,\alpha} \left(\hat{c}^{\dag}_{\alpha,i} \hat{c}_{\alpha,i+1} + \mathrm{h.c.}\right)
+ U \sum_{i, \alpha,\beta \neq \alpha} \hat{n}_{\alpha i} \hat{n}_{\beta i} + \\
& +\frac{\Omega}{2}                 \sum_{i}\left( \hat{c}^{\dag}_{1,i} \hat{c}_{2,i}  + \mathrm{h.c.}\right)+ \frac{\delta}{2} \sum_{i} \left( \hat{n}_{1,i} - \hat{n}_{2,i} \right)
\end{aligned}
\end{equation}
where we have added the last term to take into account the detuning of the Raman coupling (already introduced in the rotated-wave-approximated Hamiltonian of Eq. (\ref{Raman_Hamiltonian})), that is crucial in the state preparation protocol.

\begin{figure*}[t!]
\includegraphics[width= \linewidth]{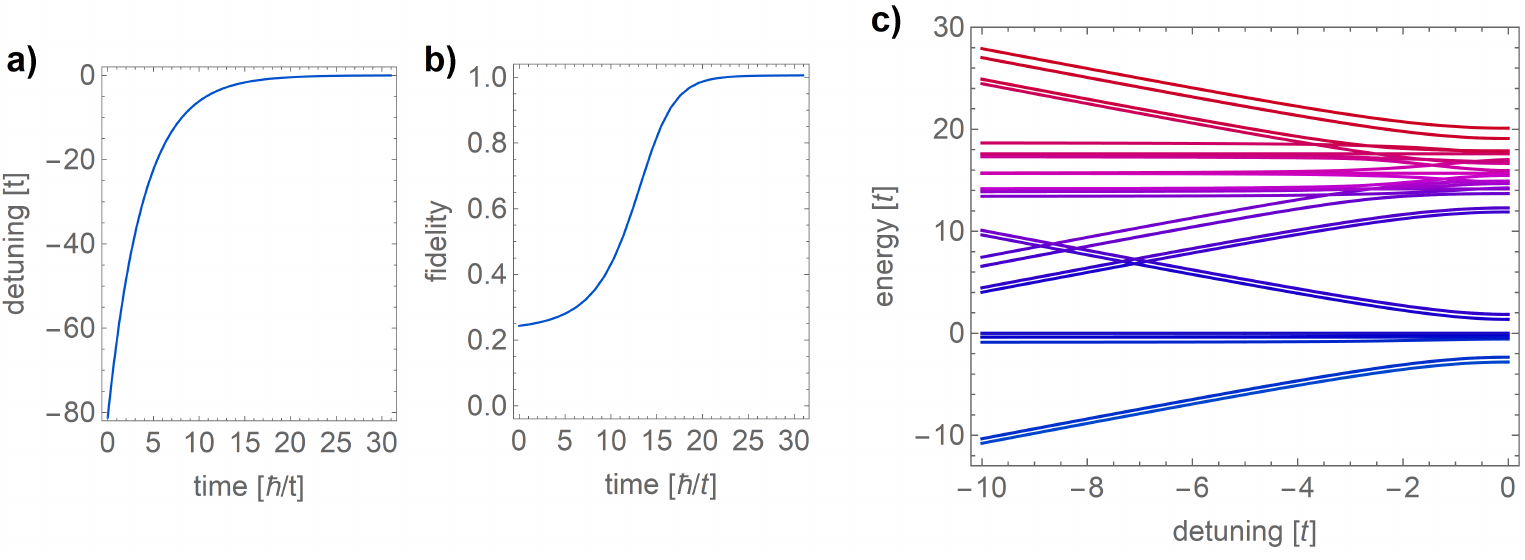}
	\caption{{\bf Numerical analysis of adiabatic state preparation.} {\bf a)} Raman detuning ramp used for the preparation of the initial state. The parameters refer to the actual ramp used in the experiment for the preparation of the system for $U=2.6D$ and $\Omega=0.35D$. {\bf b)} Calculated fidelity $|\langle\Psi_f|\Psi(t)\rangle|^2$ of the time-evolved state $|\Psi(t)\rangle$ with respect to the target ground state at the end of the ramp $|\Psi_f\rangle$. The calculation has been performed for a small system of 3 particles in 3 sites for $U=2.6D$, $\Omega=0.35D$ and the same detuning ramp shown in panel a). {\bf c)} Full spectrum of the system for $U=2.6D$ and $\Omega=0.35D$ and different Raman detunings (see text fore more details).}
	\label{fig:adiabloading}
\end{figure*}

We simulate the effect of our loading procedure for a system composed by $N=3$ particles in a lattice with $N_P=3$ sites. In this case, the dimension of the Hilbert space associated to the system is $\binom{3\,N_S}{N}=84$. Scaling the problem to bigger lattices is possible but computationally expensive. In order to simulate the loading scheme with $\Omega \ne 0$, we fix the initial populations to be $\{N_1,N_2,N_3\}=\{2,0,1\}$, where $N_{\alpha}$ refers to the total number of particles in state $\alpha$.  We determine the ground state $|\Psi_0\rangle$ before the Raman loading as the lowest-energy eigenstate of the Hamiltonian in Eq. (\ref{toy_model_hamiltonian}), with $\Omega$=0, that is compatible with the  constraint on the populations (i.e. $\{2,0,1\}$). The Raman loading is then simulated by solving the time-dependent Schr{\"o}dinger equation in the presence of all the terms of Eq. (\ref{toy_model_hamiltonian}), where $\Omega$ is kept fixed at the final value and $\delta$ is swept from $\delta \gg \Omega$ to $\delta=0$ according to Eq. (\ref{Raman_detuning}), as discussed in Sec. \ref{sec:adiabaticpreparation}.

In order to validate the loading procedure we calculate the fidelity between the time-evolved state $|\Psi(t)\rangle=e^{-i \hat{H} t/\hbar} |\Psi_0\rangle$ and the target wavefunction, that is defined as the lowest-energy eigenstate $|\Psi_f\rangle$ of the final Hamiltonian with $\delta=0$ that is compatible with the constraint on the populations $N_1+N_2=2$ and $N_3$=1. An example is shown in Fig. \ref{fig:adiabloading}b for the actual detuning ramp used in the experiment to prepare the state at $U=2.6D$ and $\Omega=0.35D$.

Fig. \ref{fig:adiabloading}c shows the full spectrum of the system as a function of the Raman detuning. There it is evident that the ground state of the system at large detuning $|\Psi_0\rangle$  is adiabatically connected with the ground state $|\Psi_f\rangle$ on resonance. We note that, even if the system initially starts with an unbalanced mixture of two species only, after this procedure it behaves as a mixtures of three species $N_1=N_2=N_3=1$ all interacting among themselves: in particular, in the presence of tunnelling and interactions between the atoms, the state rotation induced by the detuning sweep does not happen uniformly in all the sites of the lattice, making the atoms initially in state $|1\rangle$ distinguishable and therefore, interacting, at the end of the state preparation sequence.

\section{Dynamical Mean-Field Theory Calculations}

We solve the model in Eq. (\ref{eq:hamiltonian}) using Dynamical Mean-Field theory (DMFT)\cite{rev_dmft}, a non-perturbative theoretical method which maps a lattice model onto an effective impurity model. The interaction of the site with the rest of the lattice is approximated by an effective dynamical bath which is determined self-consistently.
This allows to fully capture the quantum dynamics while freezing spatial fluctuations beyond mean-field. We solve the model in a Bethe lattice of bandwidth $W$, which is known to account correctly for the physics of three-dimensional lattices. The impurity model is solved using an exact-diagonalization solver developed at SISSA\cite{capone_prbed} which requires to describe the bath in terms of a finite number of levels $N_b$ that we fix to 9 in the calculations reported in this work.

DMFT is naturally formulated in a grandcanonical ensemble, where the total density is fixed by a chemical potential $\mu$ and the occupations of the various flavours are not fixed. 
In order to {enforce the experimental population constraint} $N_1 + N_2 = 2 N_3$ in the presence of the coupling $\Omega$, we have to include an external field which counterbalances the tendency to favour (for total occupation of 1) the occupation of the coupled flavours $|1\rangle$ and $|2\rangle$. Therefore the Hamiltonian is supplemented by the terms
\begin{equation}
\hat{H}_{fields} = \sum_i \left( -\mu\sum_{\alpha} \hat{n}_{\alpha i} +h \left( \hat{n}_{1i} + \hat{n}_{2i} -2\hat{n}_{3i} \right) \right),
\label{hfields}
\end{equation}
where $\mu$ and $h$ have to be determined self-consistently to reach a total occupation of 1 fermion per site.
For a uniform infinite lattice, we have drawn a phase diagram based on the flavour-resolved quasiparticle weight $Z_{\alpha}$, which is 1 for a non-interacting fermion and it is reduced by the interactions.
The global Mott transition is reached when all the $Z_{\alpha}$ vanish, while a sharp crossover is observed between a standard metal and a region of selective correlation where the quasiparticle weight of the coupled species rapidly drops to a value smaller than 0.05.

Besides the calculations for a uniform lattice, we also have used DMFT to estimate the effect of the harmonic trapping potential
\begin{equation}
\hat{V}_T=\kappa \sum_{i,\alpha} R_i^2 \hat{n}_{\alpha i},
\label{hfields}
\end{equation}
where $R_i$ is the distance of the site i from the center of the trap.
In order to reduce the computational cost, we have employed a local-density approximation (LDA), which assumes that the local properties of the system can be computed from a uniform model with the local values of the physical parameters, in the present case of the potential given by the harmonic trap. This amounts to have a local value of the chemical potential $\mu' = \mu -\kappa \sum_{i,\alpha} R_i^2$.

In order to build the global observables for the trapped system, we have solved the model given by Eqs. (\ref{eq:hamiltonian}) and (\ref{hfields}) for a wide range of values of $\mu'$ spanning the whole range of densities from 0 to 3. For every value of $\mu'$ we have found the value of $h$ which gives $N_1 = N_2 = N_3$. 

The global observables (total number of particles, doublons, flavour-selective doublons) are then obtained integrating the local counterparts over the whole system. In this way we obtain the plots of doublons as a function of the total number of particles shown in the manuscript.

\section{Density Matrix Renormalization Group Calculations}

We also solve the model in Eq. (\ref{eq:hamiltonian}) by means of a Density Matrix Renormalization Group (DMRG) calculation powered by the ITensor library \cite{itensor}. We fix the size of the system as $L=30$ and the total number of particles as $N=21$. We simulate the internal states as effective physical sites with the redefinition of proper tunnelling and interaction couplings, which results in a total of $N_{L}=90$ lattice sites. We find the ground state of the Hamiltonian in Eq. (\ref{eq:hamiltonian}) by performing $35-50$ DMRG sweeps, with a maximum bond dimension in the interval $(20,200)$. To enforce particle number conservation in the Raman-coupled case, we introduce an additional chemical potential defined as the $h$ term in Eq. (\ref{hfields}). The number of doubly occupied sites shown in Fig. \ref{fig3} also includes the number of triple occupancies (giving a maximum contribution of $\sim 10\%$ at small values of $U$, $\Omega$).


\begin{thebibliography}{99}
	\bibitem{mott} N. F. Mott, Proceedings of the Physical Society of London Series A {\bf 62}, 416 (1949).
	\bibitem{pwa} P.W. Anderson, The resonating valence bond state in La2CuO4 and superconductivity, Science {\bf 235}, 1196 (1987).
	\bibitem{review_lee} P. A. Lee, N. Nagaosa, and X.-G. Wen, Doping a Mott insulator: Physics of high-temperature superconductivity, Rev. Mod. Phys. {\bf 78}, 17 (2006).
	\bibitem{iceberg} S. Paschen and Q. Si, Nat. Rev. Phys. {\bf 3}, 9 (2021)
	\bibitem{review_gdm} A. Georges, L. de' Medici, J. Mravlje, Strong electronic correlations from Hund's coupling, Annual Reviews of Condensed Matter Physics {\bf 4}, 137 (2013)
	\bibitem{review_vojta} M. Vojta, Orbital-selective Mott transitions: Heavy fermions and beyond, J. Low. Temp. Phys. {\bf 161}, 203 (2010)
	\bibitem{hosono} Y. Kamihara, T. Watanabe, M. Hirano, and H. Hosono, Iron-Based Layered Superconductor La[O$_{1-x}$F$_{x}$]FeAs (x=0.05--0.12) with T$_c$=26 K, Am. Chem. Soc. {\bf  130}, 3296 (2008)
	\bibitem{prl_dgc} L. de’ Medici, G. Giovannetti, and M. Capone, Selective Mott Physics as a Key to Iron Superconductors, Phys. Rev. Lett. {\bf 112}, 177001 (2014)
	\bibitem{natmat_seamus} A. Kostin, P. O. Sprau, A. Kreisel, Yi Xue Chong, A. E. Böhmer, P. C. Canfield, P. J. Hirschfeld, B. M. Andersen and J. C. Séamus Davis, Imaging orbital-selective quasiparticles in the Hund’s metal state of FeSe, Nature Materials {\bf 17}, 869 (2018)
	\bibitem{natmat_c} M. Capone, Orbital-selective metals, Nature Materials {\bf 17}, 855 (2018)
	\bibitem{seamus} P. O. Sprau, A. Kostin, A. Kreisel, A. E. Böhmer, V. Taufour, P. C. Canfield, S. Mukherjee P. J. Hirschfeld, B. M. Andersen, and J. C. Séamus Davis, Discovery of orbital-selective Cooper pairing in FeSe, Science  {\bf 357}, 75 (2017)
	\bibitem{delre2018} L. Del Re and M. Capone, Selective insulators and anomalous responses in three-component fermionic gases with broken SU(3) symmetry, Phys. Rev. A {\bf 98}, 063628 (2018).
	\bibitem{cazalilla2014} M. A. Cazalilla and A. M. Rey, Ultracold Fermi Gases with Emergent SU(N) Symmetry, Rep. Prog. Phys. {\bf 77}, 124401 (2014).
	\bibitem{gorshkov2010} A. V. Gorshkov, M. Hermele, V. Gurarie, C. Xu, P. S. Julienne, J. Ye, P. Zoller, E. Demler, M. D. Lukin, and A. M. Rey, Two-orbital SU(N) magnetism with ultracold alkaline-earth atoms, Nature Phys. {\bf 6}, 289 (2010).
	\bibitem{pagano2014} G. Pagano, M. Mancini, G. Cappellini, P. Lombardi, F. Sch{\"a}fer, H. Hu, X.-J. Liu, J. Catani, C. Sias, M. Inguscio, and L. Fallani, A one-dimensional liquid of fermions with tunable spin, Nature Phys. {\bf 10}, 198--201 (2014).
	\bibitem{taie2012} S. Taie, R. Yamazaki, S. Sugawa, and Y. Takahashi, An SU(6) Mott insulator of an atomic Fermi gas realized by large-spin Pomeranchuk cooling, Nature Phys. {\bf 8}, 825--830 (2012).
	\bibitem{hofrichter2016} C. Hofrichter, L. Riegger, F. Scazza, M. H{\"o}fer, D. Rio Fernandes, I.Bloch, and S. F{\"o}lling, Direct Probing of the Mott Crossover in the SU(N) Fermi-Hubbard Model, Phys. Rev. X {\bf 6}, 021030 (2016).
	\bibitem{ozawa2018} H. Ozawa, S. Taie, Y. Takasu, and Y. Takahashi, Antiferromagnetic Spin Correlation of SU(N) Fermi Gas in an Optical Superlattice, Phys. Rev. Lett. {\bf 121}, 225303 (2018).
	\bibitem{jordens2008} R. J{\"o}rdens, N. Strohmaier, K. G{\"u}nter, H. Moritz, and T. Esslinger, A Mott insulator of fermionic atoms in an optical lattice, Nature {\bf 455}, 204--207 (2008).
	\bibitem{supplementary} See Supplementary Materials.
	\bibitem{jordens2010} R. J{\"o}rdens, L. Tarruell, D. Greif, T. Uehlinger, N. Strohmaier, H. Moritz, T. Esslinger, L. De Leo, C. Kollath, A. Georges, V. Scarola, L. Pollet, E. Burovski, E. Kozik, and M. Troyer, Quantitative Determination of Temperature in the Approach to Magnetic Order of Ultracold Fermions in an Optical Lattice, Phys. Rev. Lett. {\bf 104}, 180401 (2010).
	\bibitem{rev_dmft} A. Georges, G. Kotliar, W. Krauth, and M. J. Rozenberg, Dynamical mean-field theory of strongly correlated fermion systems and the limit of infinite dimensions, Rev. Mod. Phys. {\bf 68}, 13 (1996).
	\bibitem{BHZ2006}B. A. Bernevig, T. L. Hughes, and S.-C. Zhang, Quantum Spin Hall Effect and Topological Phase Transition in HgTe Quantum Wells, Science {\bf 314}, 5806 (2006).
	\bibitem{osmt2009}L. de’ Medici, S. R. Hassan, M. Capone, and X. Dai, Orbital-Selective Mott Transition out of Band Degeneracy Lifting, Phys. Rev. Lett. {\bf 102}, 126401 (2009).
	\bibitem{barbarino2015}S. Barbarino, L. Taddia, D. Rossini, L. Mazza, and R. Fazio, Magnetic crystals and helical liquids in alkaline-earth fermionic gases, Nat. Comm. {\bf 6}, 8134 (2015).
	\bibitem{calvanese2017}M. Calvanese Strinati, E. Cornfeld, D. Rossini, S. Barbarino, M. Dalmonte, R. Fazio, E. Sela, and L. Mazza, Laughlin-like States in Bosonic and Fermionic Atomic Synthetic Ladders, Phys. Rev. X {\bf 7}, 021033 (2017).
	\bibitem{amaricci2015}A. Amaricci, J. C. Budich, M. Capone, B. Trauzettel, and G. Sangiovanni, First-Order Character and Observable Signatures of Topological Quantum Phase Transitions, Phys. Rev. Lett. {\bf 114}, 185701 (2015)
	\bibitem{amaricci2017}A. Amaricci, L. Privitera, F. Petocchi, M. Capone, G. Sangiovanni, and B. Trauzettel, Edge state reconstruction from strong correlations in quantum spin Hall insulators, Phys. Rev. B {\bf 95}, 205120 (2017)
\end{thebibliography}

\begin{thebibliography}{99}

    \bibitem[S1]{Han2018} J. H. Han, J. H. Kang, M. Lee, and Y. Shin,  Photoassociation spectroscopy of ultracold $^{173}$Yb atoms near the intercombination line, Phys. Rev. A {\bf 97}, 013401 (2018). 
    \bibitem[s2]{andrewNG} A. Coates and A. Y. Ng, Learning Feature Representations with K-means, {\it Neural Networks: Tricks of the Trade} (Springer, 2012).
    \bibitem[S3]{Mancini2015} M. Mancini, G. Pagano, G. Cappellini, L. Livi, M. Rider, J. Catani, C. Sias, P. Zoller, M. Inguscio, M. Dalmonte, and L. Fallani, Observation of chiral edge states with neutral fermions in synthetic Hall ribbons, Science {\bf 349}, 1510 (2015). 
    \bibitem[S4]{Mikio2017} M. Miranda, R. Inoue, N. Tambo, and M. Kozuma, Site-resolved imaging of a bosonic Mott insulator using ytterbium atoms, Phys. Rev. A {\bf 96}, 4, 043626 (2017).
    \bibitem[S5]{capone_prbed} M. Capone, L. de’ Medici, and A. Georges, Solving the dynamical mean-field theory at very low temperatures using the Lanczos exact diagonalization, Phys. Rev. B {\bf 76}, 245116 (2007).
    \bibitem[S6]{itensor} \mbox{ITensor Library} (version 3.1) http://itensor.org

\end{thebibliography}
\end{document}